\documentclass[runningheads]{llncs}
\usepackage{graphicx}
\usepackage[hidelinks]{hyperref}
\usepackage{tikz}
\usepackage{amsmath,bm}
\usepackage{dsfont}
\usepackage{siunitx}
\usepackage{algorithm2e}

\PassOptionsToPackage{hyphens}{url}\usepackage{hyperref}
\usepackage{amssymb}
\usepackage{todonotes}
\usepackage{float}
\usepackage{booktabs}
\usepackage{xcolor}
\usepackage{multirow}
\usepackage{multicol}
\usepackage{booktabs}
\usepackage{color, colortbl}
\usepackage{listings}

\definecolor{darkred}{RGB}{193, 39, 45}
\definecolor{indigo}{RGB}{0, 0, 167}
\definecolor{teal}{RGB}{0, 129, 118}
\definecolor{yellow}{RGB}{238, 204, 22}
\definecolor{lightgray}{RGB}{179, 179, 179}

\begin{document}
\title{Parallel FFTW on RISC-V: A Comparative Study including OpenMP, MPI, and HPX}
\titlerunning{Parallel FFTW on RISC-V}
\author{Alexander Strack\inst{1}\orcidID{0000-0002-9939-9044}  \and Christopher Taylor \inst{2}\orcidID{0000-0001-7119-818X} 
\and Dirk Pflüger\inst{1} \orcidID{0000-0002-4360-0212}}
\authorrunning{A. Strack \and et. al}
\institute{Institute of Parallel and Distributed Systems, University of Stuttgart,\\ 70569 Stuttgart, Germany\\ \email{\{alexander.strack, dirk.pflueger\}@ipvs.uni-stuttgart.de}
\and Tactical Computing Labs LLC, 1001 Pecan St., Lindsay, Texas, U.S.A.
\email{ctaylor@tactcomplabs.com}
}

\maketitle              % typeset the header of the contribution
\begin{abstract}
Rapid advancements in RISC-V hardware development shift the focus from low-level optimizations to higher-level parallelization. 
Recent RISC-V processors, such as the SOPHON SG2042, have 64 cores. RISC-V processors with core counts comparable to the SG2042, make efficient parallelization as crucial for RISC-V as the more established processors such as x86-64. 

In this work, we evaluate the parallel scaling of the widely used FFTW library on RISC-V for MPI and OpenMP. We compare it to a 64-core AMD EPYC 7742 CPU side by side for different types of FFTW planning.
Additionally, we investigate the effect of memory optimization on RISC-V in HPX-FFT, a parallel FFT library based on the asynchronous many-task runtime HPX using an FFTW backend.

We generally observe a performance delta between the x86-64 and RISC-V chips of factor eight for double-precision 2D FFT. Effective memory optimizations in HPX-FFT on x86-64 do not translate to the RISC-V chip. 
%Using jemalloc with HPX on RISC-V can results in significant performance improvements. 
FFTW with MPI shows good scaling up to 64 cores on x86-64 and RISC-V regardless of planning. In contrast, FFTW with OpenMP requires measured planning on both architectures to achieve good scaling up to 64 cores.

The results of our study mark an early step on the journey to large-scale parallel applications running on RISC-V.
\keywords{FFTW \and RISC-V \and OpenMP \and MPI \and HPX \and FFT}
\end{abstract}
\setcounter{footnote}{0} 
\section{Introduction}\label{sec:introduction}
The RISC-V \cite{Waterman2014_riscv} hardware architecture emerges as a promising alternative to established architectures such as x86-64 or ARM, both of which are locked behind license.
This trend is supported by the EuroHPC \footnote{\url{https://eurohpc-ju.europa.eu/index_en}(accessed on 05/16/2025)} that actively promotes the development of new RISC-V hardware. 
As a result, new projects like the \textit{European Processor initiative} \footnote{\url{https://www.european-processor-initiative.eu/project/epi/}(accessed on 05/16/2025)} or the \textit{Digital Autonomy with RISC-V in Europe} project were created. Their objective extends beyond the development of RISC-V hardware, to support the complete software stack of existing high-performance computing (HPC) applications such as large-scale simulations or artificial intelligence workloads.

One of the essential HPC software building blocks is the fast Fourier transform (FFT). 
The Fourier transform is a fundamental tool for converting data from the spatial domain into the frequency domain. FFTs accelerate convolution operations and serve as an integral part of many scientific applications such as in molecular dynamics \cite{Deserno1998}.
Undoubtedly, the most popular FFT library is the open source project \textit{Fastest Fourier Transform in the West} (FFTW).
More recent libraries show minor improvements in parallel performance \cite{Ayala2022}. However, FFTW serves as a base for many of these newer parallel FFT libraries.

In this work, we evaluate the performance of parallel FFTW on the SG2042 64-core RISC-V chip recently developed by SOPHGO \footnote{\url{https://en.sophgo.com/sophon-u/product/introduce/sg2042.html}\newline(accessed on 05/16/2025)}.
We compare the performance of different parallelization approaches side-by-side with a 64-core AMD EPYC 7742 \footnote{\url{https://www.amd.com/en/support/downloads/drivers.html/processors/epyc/epyc-7002-series/amd-epyc-7742.html} (accessed on 05/16/2025)} chip based on the x86-64 architecture. 
Based on our previous work \cite{Strack_2024}, we also benchmark the HPX-FFT tool and investigate the impact of different memory access optimizations.
Furthermore, we also compare HPX-FFT with parallel FFTW for different planning types.

Our main contributions in this work include, to the best of our knowledge:
\begin{itemize}
    \item The first side-by-side comparison of parallel FFT on x86-64 and RISC-V hardware.
    \item The first evaluation of multidimensional FFTW planning on up to 64 RISC-V cores.
    \item The first comparison of asynchronous tasking with HPX and MPI+OpenMP on RISC-V.
\end{itemize}

The remainder of this work has the following structure:
In Section \ref{sec:related_work}, we review related work on FFT and HPX on RISC-V.
Then in Section \ref{sec:framework}, we introduce the FFTW library and give an overview of the different parallelization approaches used in FFTW and HPX-FFT. 
In Section \ref{sec:methods}, we explain the basics of parallel multidimensional FFT and describe our benchmark setup.
The results of our side-by-side comparison of RISC-V and x86-64 are discussed in Section \ref{sec:results}.
Lastly, in Section \ref{sec:conclusion} we conclude our work and give an outlook on future work regarding FFTW on RISC-V and other architectures.

\section{Related work}\label{sec:related_work}
Several studies focused on FFT implementations targeting RISC-V. The studies assess FFTW as a reference point for performance comparison.
In \cite{Vizcaino2021_1d_fftw_riscv}, the authors use a scalar FFTW to compare performance against their vectorized approaches on a RISC-V core prototype. The authors of \cite{Li2023_1d_fftw_risv}, benchmark their FFTASI framework against FFTW on a C910MP chip.
In \cite{Zhao2023_1d_fftw_riscv}, the authors benchmark their PerfMPL-FFT library against FFTW in single and double precision on a C910MP chip. 
Those works are motivated by FFTW's singular support for the ratified RVV1.0 specification.

Many of the current RISC-V chips only support the unratified RVV0.7.1 specification. The author of \cite{Cai2024_simd_fftw_riscv} implemented RVV0.7.1 support for FFTW and compared the performance on several RISC-V chips. All these works on FFTW with RISC-V have one thing in common: FFTW is always used sequentially to compute a 1D FFT.  In contrast, we investigate parallel 2D FFT with FFTW.

The first efforts to port the HPX runtime to RISC-V were made in \cite{Diehl2023_riscv}. 
The authors evaluated the parallel performance and energy consumption of different hardware architectures using the Octo-Tiger astrophysics library \cite{Marcello21_octotiger}.
In \cite{Diehl2024_riscv} their RISC-V work is extended to benchmark vectorization with Octo-Tiger on the SG2042 chip.
A detailed performance analysis and comparison of the SG2042 chip against x86-64 is given in \cite{Brown_2023}. 
Their comparison includes the two chips used in our work. When applied to a double-precision workload, the EPYC 7742 chip out performed the SG2042 chip by a factor of five. 
The authors use the RAJA performance suite \cite{Hornung2017_raja} to benchmark the system for different workloads. However, the RAJA performance suite does not include the FFT algorithm and MPI-based parallelization.

\section{Software stack}\label{sec:framework}
In the following subsections, we discuss the idea behind the FFTW \cite{Frigo2005} library and its different parallelization approaches. Furthermore, we introduce HPX-FFT and the asynchronous many-task runtime HPX.

\subsection{FFTW}
FFTW is the de facto standard for FFT in HPC. 
To date, its hardware adaptivity and vectorization make FFTW one of the fastest libraries to compute multidimensional FFT. 
Although in distributed memory environments, libraries like P3DFFT \cite{Pekurovsky2012}, exhibit a slight performance advantage due to a more advanced data distribution strategy. 
One major drawback of FFTW is its lack of accelerator support.
Developers seeking to benefit from accelerators are required to incorporate hybrid solutions themselves or rely on libraries such as AccFFT \cite{Gholami2015} that combine FFTW with GPU FFT libraries.

The compilation of FFTW is tedious and not recommended for end-users.
Inside this tedious process lies the secret of FFTW's performance. 
FFTW uses code generation to produce highly optimized FFT codelets that serve as a basis for large-scale transformations. 
The best-performing combination of these codelets is different from system to system. 
In FFTW, the art of finding the optimal combination is called \textit{planning}. 
For planning, FFTW performs an offline test on a set of combinations for a FFT of given size and dimension. 
The result is called a \textit{plan} and can be reused multiple times.
This makes it perfect for iterative workloads typically found in scientific computing applications. 
FFTW can rely on precomputed FFT plans in the form of \textit{wisdom}. FFTW's \textit{wisdom} is a cache of optimal performance parameters used in an FFT execution plan.
The FFTW library supports multiple types of planning. We focus on \textit{estimated} and \textit{measured} planning. 
Estimated planning uses an exclusive heuristic to approximate the best combination of codelets. In contrast, measured planning uses a reduced subset of parameter combinations.

FFTW supports three different parallelization variants out-of-the-box: MPI, OpenMP, and POSIX threads. 
We omit the HPX backend developed in \cite{Strack_2024} from this list as it is not yet part of an official FFTW release. The HPX backend requires compiling FFTW from source. In the following subsections, we discuss the MPI and OpenMP backends in more detail. The POSIX backend is not considered in this work because of implementation and performance similarities with the OpenMP backend.

\subsection{OpenMP}
The Open Multiprocessing Standard \cite{Dagum1998_openmp} was added to C/C++ in 1999. 
It contains a set of compiler directives that allow for easy-to-implement shared memory parallelization. 
The main parallelization concept of OpenMP is built on the fork-join principle.
As an example, when the compiler encounters \texttt{\#pragma omp parallel for} to schedule an embarrassingly parallel for loop, the iteration range of the loop is partitioned. Each partition forks threads.
Each thread works on a different part of the loop. 
After all threads have finished their work, they are joined back together. 
The remaining portions of the application are executed sequentially until the next directive. 
Note that since OpenMP 3.0 tasking is also supported.

In FFTW, the OpenMP backend optimizes the critical path and number of threads for minimal parallelization overhead.
The OpenMP backend can be easily incorporated into existing sequential FFTW code. Threading is initialized by calling the \texttt{fftw\_init\_threads()} function.
For multithreading-aware planning invoke \texttt{fftw\_plan\_with\_threads(n\_threads)}.
Here \texttt{n\_threads} resembles the number of OpenMP threads.

\subsection{MPI}
Since its release in 1994, the Message Passing Interface \cite{Mpi2012} has evolved into the standard for communication between distributed physical or logical memory entities, referred to as ranks in MPI terminology.
As the name suggests, MPI is based on sending and receiving data packed into messages. MPI supports both synchronous (blocking) and asynchronous (non-blocking) point-to-point communication.  
For complex parallel algorithms, point-to-point communication quickly becomes tedious and error-prone. 
MPI tackles the complexity by providing a set of pre-defined communication patterns in the form of collective operations. 
Collectives operate on participating ranks grouped within a communicator.
MPI only serves as a standard that defines syntax and functionalities. 
There exist multiple implementations of this standard such as OpenMPI \cite{Gabriel2004_openmpi} and MPICH \cite{Groop1999_mpich}.

Parallel multidimensional FFT communication follows a clear pattern. It perfectly matches the all-to-all collective pattern. 
FFTW uses \texttt{MPI\_all\_to\_all} to handle communication between ranks.
Using the MPI backend in FFTW requires including the MPI-specific header file.
The MPI backend is initialized with a call to \texttt{fftw\_mpi\_init()}.
Dimension-specific \texttt{fftw\_mpi\_local\_size\_<dim>d} functions then determine the local data size on each rank in an MPI communicator.
Special \texttt{fftw\_mpi\_plan\_<type>\_<dim>d} functions enable planning on data distributed across multiple ranks. 
It is also possible to combine the MPI with the OpenMP backend. As a result, we can run FFTW with MPI+OpenMP. Note that \texttt{MPI\_THREAD\_FUNNELED} has to be set for this combination to work correctly. 
For a more detailed explanation of how to run FFTW in parallel, we refer to the FFTW3 documentation for shared\footnote{\url{https://www.fftw.org/fftw3_doc/Multi_002dthreaded-FFTW.html\#Multi_002dthreaded-FFTW} (accessed on 05/16/2025)} and distributed\footnote{\url{https://www.fftw.org/fftw3_doc/Distributed_002dmemory-FFTW-with-MPI.html\#Distributed_002dmemory-FFTW-with-MPI} (accessed on 05/16/2025)} tutorials.

\subsection{HPX-FFT}
HPX-FFT is a software tool developed to compute 2D FFTs utilizing the asynchronous many-task runtime HPX.
It is targeted towards shared-memory and distributed environments of arbitrary scale. 
HPX-FFT is built on HPX \cite{kaiser2020hpx}, the standard C++ library for concurrency and parallelism, to enable the parallelization of 1D FFTs.
FFTW is used as the backend for computing the 1D FFTs. 
For a more detailed explanation of the parallelization approach, see Section~\ref{sec:fft}.

HPX facilitates asynchronous parallelization through the use of futures to define task dependencies. Task dependencies can be chained together creating
a task graph. Task graphs are efficiently scheduled by work-stealing schedulers and executed via lightweight HPX threads. 
HPX provides an Active Global Address Space (AGAS) for communication. 
The AGAS enables both implicit and explicit communication via active messages. 
These active messages, named \textit{parcels}, are sent over a so-called \textit{parcelport}.
At the time of writing, HPX supports three parcelports: TCP, MPI, and LCI \cite{Yan23}. A performance comparison of these parcelports using HPX-FFT is given in \cite{Strack2025_hpxfft}.

In this work, we use HPX-FFT as a reference point to assess the quality of parallel 2D FFTW plans. Furthermore, we use HPX-FFT as an example of parallel memory access pattern optimization.

\section{Methods}\label{sec:methods}
In this section, we briefly introduce the Fourier transform and its extension to multiple dimensions. Furthermore, we present a robust parallelization scheme. 
In the end, we state the different implementations of HPX-FFT and FFTW used in our benchmarks.

\subsection{Parallel FFT}\label{sec:fft}
For a real-valued discrete 1D signal $f=[f_0,....,f_{N-1}]$, the discrete Fourier transform is given by
\begin{equation}\label{eq:ft}
    \hat{f}_k = \sum^{N-1}_{n=0} f_n \cdot \phi_k(n)   
\end{equation} for $k=0, ...,\frac{N}{2}$.
With different basis functions $\phi_k(n)$ we can perform Fourier, sine, and cosine transforms.
%FFT
To expand the Fourier transform into multiple dimensions, a 1D transform is performed sequentially in every dimension.
For a multidimensional signal containing real values, this means that the first transform changes from real to complex space (r2c). 
All subsequent transforms remain in the complex space (c2c).
Assuming the signal is stored in row-major storage format, only the first dimension is stored contiguously in matrix rows.
Thus, for a 2D Fourier transform, there are two options to handle the second 1D transform. 
Either the transform is computed with index offsets or the matrix is transposed to compute the second dimension in contiguous memory.
If the signal is small, the first approach is typically faster. However, as soon as the matrix rows do not fit in the cache anymore, the second approach is favorable. 

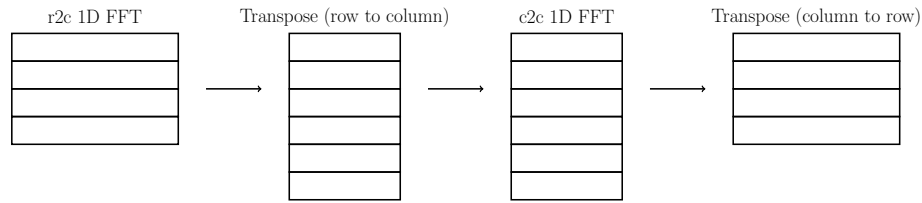
\begin{figure}[h]
    \centering
    \resizebox{1\textwidth}{!}{
    \begin{tikzpicture}
        \tikzstyle{every node}=[font=\Huge]
        \draw [ line width=2pt ] (-8.75,20) rectangle (-1.25,18.75);
        \draw [ line width=2pt ] (-8.75,18.75) rectangle (-1.25,17.5);
        \draw [ line width=2pt ] (-8.75,17.5) rectangle (-1.25,16.25);
        \draw [ line width=2pt ] (-8.75,16.25) rectangle (-1.25,15);

        \draw [ line width=2pt ] (3.75,20) rectangle (8.75,18.75);
        \draw [ line width=2pt ] (3.75,18.75) rectangle (8.75,17.5);
        \draw [ line width=2pt ] (3.75,17.5) rectangle (8.75,16.25);
        \draw [ line width=2pt ] (3.75,16.25) rectangle (8.75,15);
        \draw [ line width=2pt ] (3.75,15) rectangle (8.75,13.75);
        \draw [ line width=2pt ] (3.75,13.75) rectangle (8.75,12.5);

        \draw [ line width=2pt ] (13.75,20) rectangle (18.75,18.75);
        \draw [ line width=2pt ] (13.75,18.75) rectangle (18.75,17.5);
        \draw [ line width=2pt ] (13.75,17.5) rectangle (18.75,16.25);
        \draw [ line width=2pt ] (13.75,16.25) rectangle (18.75,15);
        \draw [ line width=2pt ] (13.75,15) rectangle (18.75,13.75);
        \draw [ line width=2pt ] (13.75,13.75) rectangle (18.75,12.5);
         
        \draw [ line width=2pt ] (23.75,20) rectangle (31.25,18.75);
        \draw [ line width=2pt ] (23.75,18.75) rectangle (31.25,17.5);
        \draw [ line width=2pt ] (23.75,17.5) rectangle (31.25,16.25);
        \draw [ line width=2pt ] (23.75,16.25) rectangle (31.25,15);

        \draw [line width=2pt, ->] (0,17.5) -- (2.5,17.5);
        \draw [line width=2pt, ->] (10,17.5) -- (12.5,17.5);
        \draw [line width=2pt, ->] (20,17.5) -- (22.5,17.5);

        \node [font=\Huge] at (-5,20.75) {r2c 1D FFT};
        \node [font=\Huge] at (6.25,20.75) {Transpose (row to column)};
        \node [font=\Huge] at (16.25,20.75) {c2c 1D FFT};
        \node [font=\Huge] at (27.5,20.75) {Transpose (column to row)};
        \end{tikzpicture}
    }%
\label{fig:parallel_fft}
\caption{Four steps of parallel 2D FFT}
\end{figure}

The naive Fourier transform has a computational complexity of $\mathcal{O}(N^{d+1})$ for $d$ dimensions. 
In 1965, Cooley and Turkey proposed the FFT algorithm \cite{Cooley1965}. 
FFT reduces the complexity of the multidimensional Fourier transform to $\mathcal{O}(N^{d}\cdot \log(N))$.
The resulting parallel algorithm using FFT and data transposition consists of four sequential steps for two dimensions (see Figure \ref{fig:parallel_fft}).
In the next subsection, we introduce different HPX-FFT implementations that all perform these four steps but differ in synchronization.

\subsection{Implementations}\label{sec:implementations}
In this work, we benchmark several different implementations that perform a 2D FFT in parallel.
To investigate the effect of memory optimization on RISC-V, we consider the following implementations:
\begin{itemize}
    \item HPX-FFT \textit{naive}: Future-based with minimal synchronization.
    \item HPX-FFT \textit{opt}: Future-based with memory access optimization.
    \item HPX-FFT \textit{sync}: Future-based with maximal synchronization.
    \item HPX-FFT \textit{for\_loop}: Fork-join-based implementation using \newline \texttt{hpx::experimental::for\_loop}.
\end{itemize}
To compare the FFTW backends, we consider the following implementations:
\begin{itemize}
    \item FFTW MPI: Parallelization with MPI ranks.
    \item FFTW OpenMP: Parallelization with OpenMP threads.
    \item FFTW MPI+OpenMP: Parallelization with a combination of OpenMP threads and MPI ranks.
\end{itemize}

\section{Results}\label{sec:results}
We divide the results into three different parts. 
First, we compare the different HPX-FFT implementations to investigate the impact of memory access optimizations on RISC-V. 
Then, we perform a scaling test of parallel FFTW using estimated planning. 
Followed by a scaling test using measured planning.

As a uniform benchmark, we choose a 2D FFT of size $2^{14}\times2^{14}$ in FP64 precision and perform strong scaling benchmarks from 1 to 64 cores on both chips.
All runtimes presented in this section are the median of ten runs.
The error bars visualize the minimum and maximum of these runs.
The results of our previous work are not reused.
Instead, we rerun the benchmarks on the x86-64 system with software versions that match the RISC-V system.

The specifications of the systems used in this work are presented in Table \ref{tab:specs}.
Note that the $64$ cores on the EPYC $7742$ chip are organized in eight clusters with eight Zen2 cores each.
Each core has $96$KB of L1 cache and $512$KB of L2 cache.
In contrast, the $64$ cores on the SG2042 chip are organized into 16 clusters of 4 RISC-V cores per cluster. Each cluster has $64$KB of data cache and $64$KB of instruction cache. Additionally, each cluster has $1$MB of L2 cache.

\begin{table}[h]
        \centering
                \resizebox{0.7\columnwidth}{!}{%
                \begin{tabular}{lll}
                    \toprule
                    Architecture         & x86-64              & RISC-V                \\ \midrule
                    %\rowcolor{lightgray}
                    CPU                  & AMD EPYC 7742    & SOPHON SG2042\\ %\hline
                    \rowcolor{lightgray!50}
                    Cores               & 64                & 64                            \\ %\hline
                    %\rowcolor{lightgray}
                    Base clock          & 2.25GHz           & 2.00GHz                    \\ %\hline
                    \rowcolor{lightgray!50}
                    L3 Cache            & 256MB             & 64MB        \\ %\hline
                    %\rowcolor{lightgray}
                    SIMD                & AVX2              & RVV0.7.1        \\ %\hline
                    \rowcolor{lightgray!50}
                    TDP                 & 225W              & 120W                     \\ \bottomrule
                \end{tabular}}
                \caption{Hardware specification of the x86-64 and RISC-V chips}
                \label{tab:specs}
\end{table}

% The SG2042 has a clock speed of 2GHz and a design consisting of 16 clusters, each containing 4 RISC-V cores, with each core featuring L1-D 64KB and L1-I 64KB. Each cluster shares a design of L2 1MB. The L3 System cache has a capacity of 64MB.
%The memory system in hardware has limitations with how many loads and stores it is capable of supporting. This creates a bottleneck that impacts performance - changing access patterns can only improve performance if it means decreasing the number of memory operations.
\subsection{Memory access optimization}
\begin{figure}
    \centering
    \begin{minipage}[t]{.47\textwidth}
        \centering
        %\raggedright
        \includegraphics[width=\linewidth]{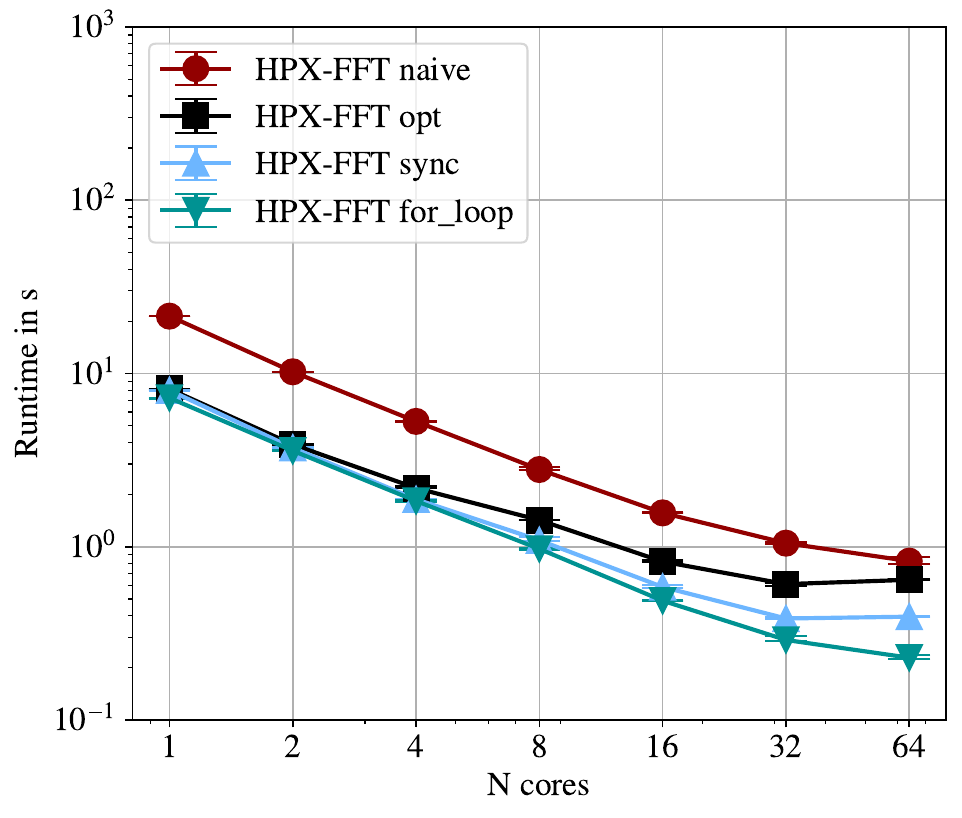}
        \caption{Strong scaling runtimes of different optimizations on up to $64$ x86-64 cores for a $2^{14} \times 2^{14}$ FFT using HPX-FFT.}
        \label{fig:hpx_x86}
    \end{minipage}\hspace{.05\textwidth}
    \begin{minipage}[t]{.47\textwidth}
        \centering
        %\raggedleft
        \includegraphics[width=\linewidth]{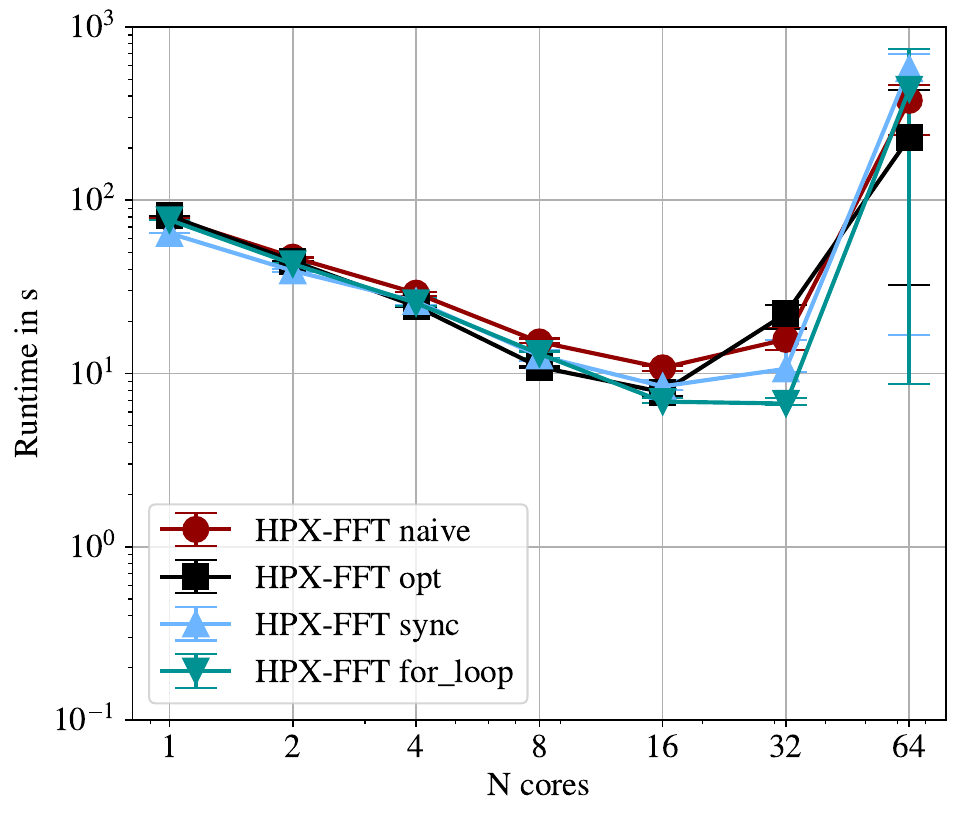}
        \caption{Strong scaling runtimes of different optimizations on up to $64$ RISC-V cores for a $2^{14} \times 2^{14}$ FFT using HPX-FFT.} 
        \label{fig:hpx_riscv}
    \end{minipage}
\end{figure}

Accordingly to \cite{Strack_2024}, we observe a significant performance difference between the different optimization levels. 
In Figure \ref{fig:hpx_x86}, the naive implementation shows a nearly constant multiplicative overhead compared to the best implementation from one core to 64 cores. 
This overhead is not significant on the RISC-V machine (see Figure \ref{fig:hpx_riscv}). 
In general, the optimizations have little impact on the RISC-V machine.
Up to 16 cores, the performance delta between RISC-V and x86-64 is around factor $5.6$ for the naive implementation and around factor $10.7$ for the remaining three implementations.
For more than 32 cores, the performance of all implementations is significantly reduced.

\begin{figure}
    \centering
    \begin{minipage}[t]{.47\textwidth}
        \centering
        %\raggedright
        \includegraphics[width=\linewidth]{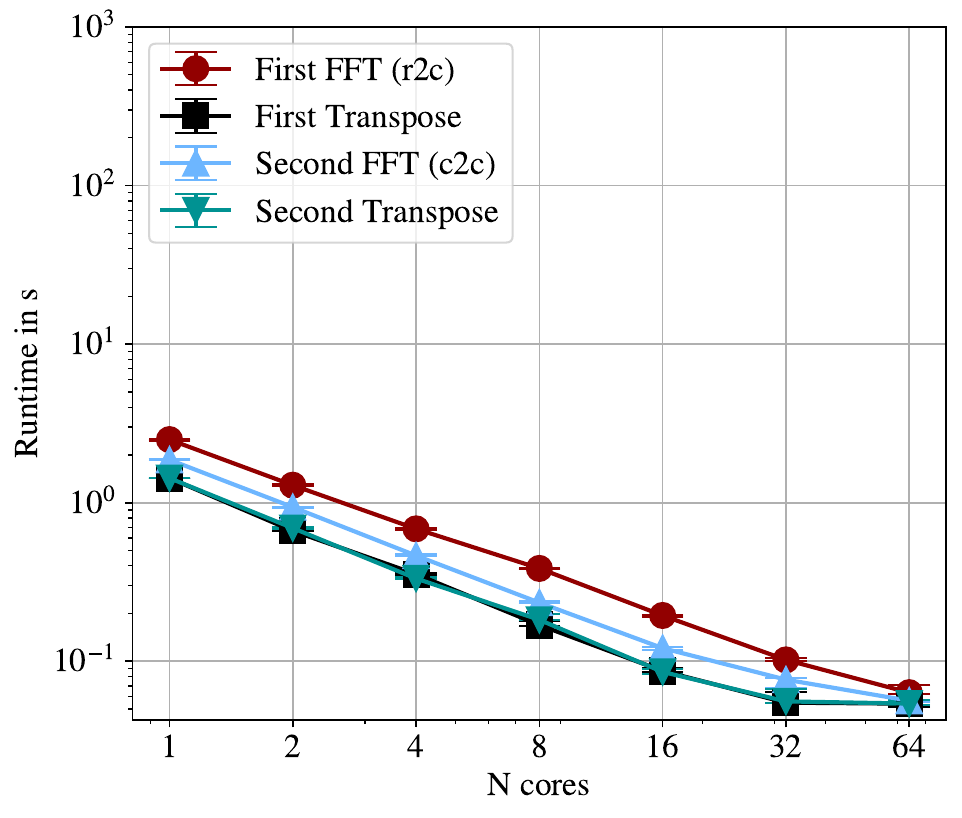}
        \caption{Partial runtimes of HPX-FFT \emph{for\_loop} on up to $64$ x86-64 cores for a $2^{14} \times 2^{14}$ FFT.}
        \label{fig:hpx_x86_part}
    \end{minipage}\hspace{.05\textwidth}
    \begin{minipage}[t]{.47\textwidth}
        \centering
        %\raggedleft
        \includegraphics[width=\linewidth]{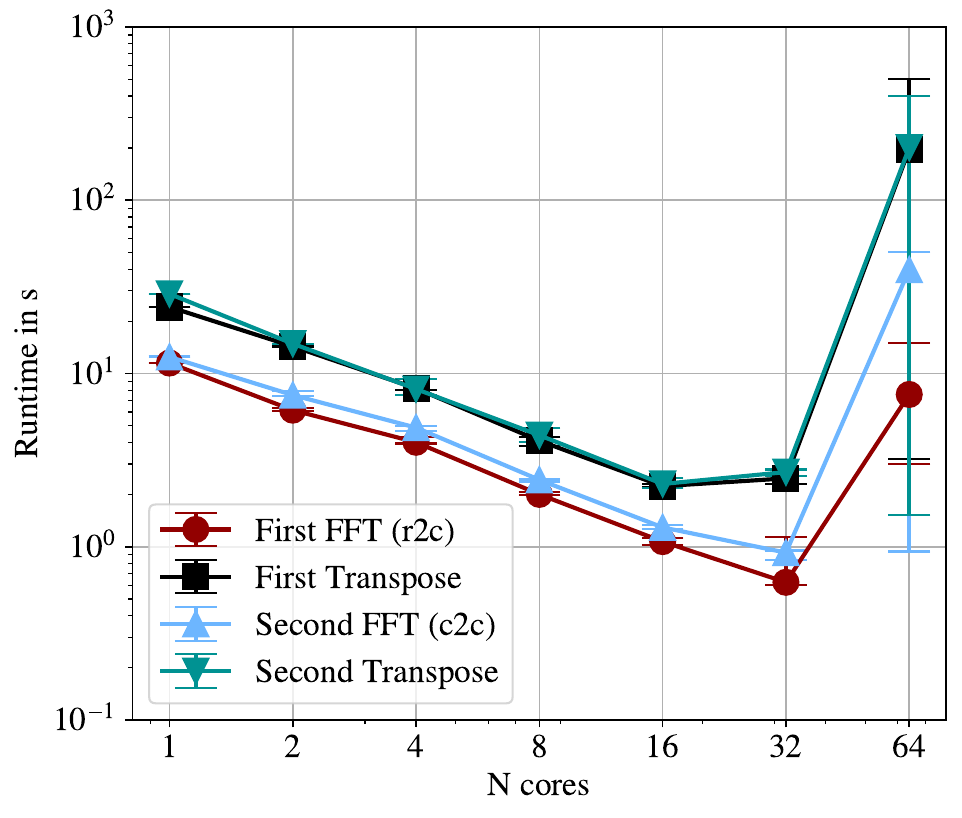}
       \caption{Partial runtimes of HPX-FFT \emph{for\_loop} on up to $64$ RISC-V cores for a $2^{14} \times 2^{14}$ FFT.}
        \label{fig:hpx_riscv_part}
    \end{minipage}
\end{figure}

Upon closer analysis of the HPX-FFT \emph{for\_loop} variant, performance bottlenecks become evident (see Figures~\ref{fig:hpx_x86_part} and~\ref{fig:hpx_riscv_part}).
On x86-64, the runtime distribution between FFT computation and data transposition is approximately 60\% to 40\% up to 32 cores and reaches 50\% to 50\% at 64 cores. The RISC-V processor contrasts as the ratios change significantly: 30\% to 70\% up to 16 cores, 20\% to 80\% at 32 cores, and 10\% to 90\% at 64 cores. This suggests less efficient memory access and cache utilization on the SG2042 chip.
%After switching from reading data from non-contiguous memory and writing to contiguous memory to the reverse approach, we observed a significant performance improvement on 32 cores. However, these changes did not lead to runtime improvements at 64 cores.

\subsection{Estimated planning}

RISC-V support for FFTW is currently under active development.
We use the most recent release on GitHub\footnote{\url{https://github.com/rdolbeau/fftw3/releases/tag/r5v-test-release-005} (accessed on 05/16/2025)} available at the time of writing.
Since the release supports RVV1.0 and the SG2042 chip only supports RVV0.7.1, we do not consider vectorization for better comparability. 

\begin{figure}
    \centering
    \begin{minipage}[t]{.47\textwidth}
        \centering
        %\raggedright
        \includegraphics[width=\linewidth]{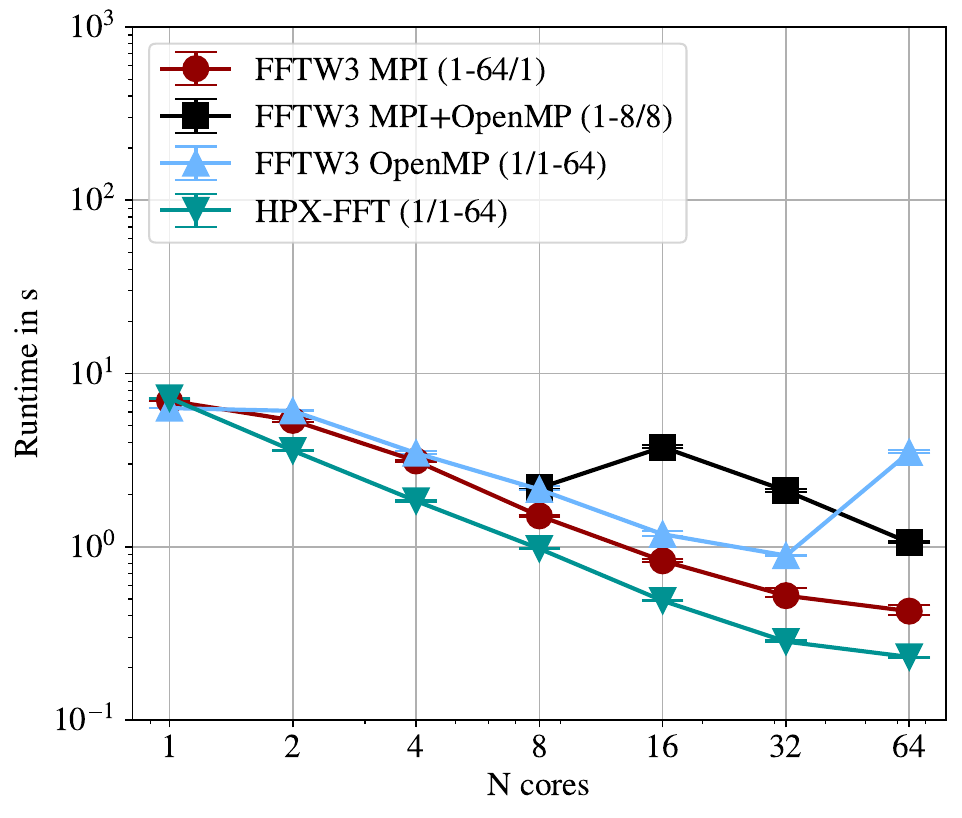}
        \caption{Strong scaling runtimes on up to $64$ x86-64 cores for a $2^{14} \times 2^{14}$ FFT using FFTW with estimated planning. The execution configuration is represented as (\#processes/\#threads).}
        \label{fig:estimate_x86}
    \end{minipage}\hspace{.05\textwidth}
    \begin{minipage}[t]{.47\textwidth}
        \centering
        %\raggedleft
        \includegraphics[width=\linewidth]{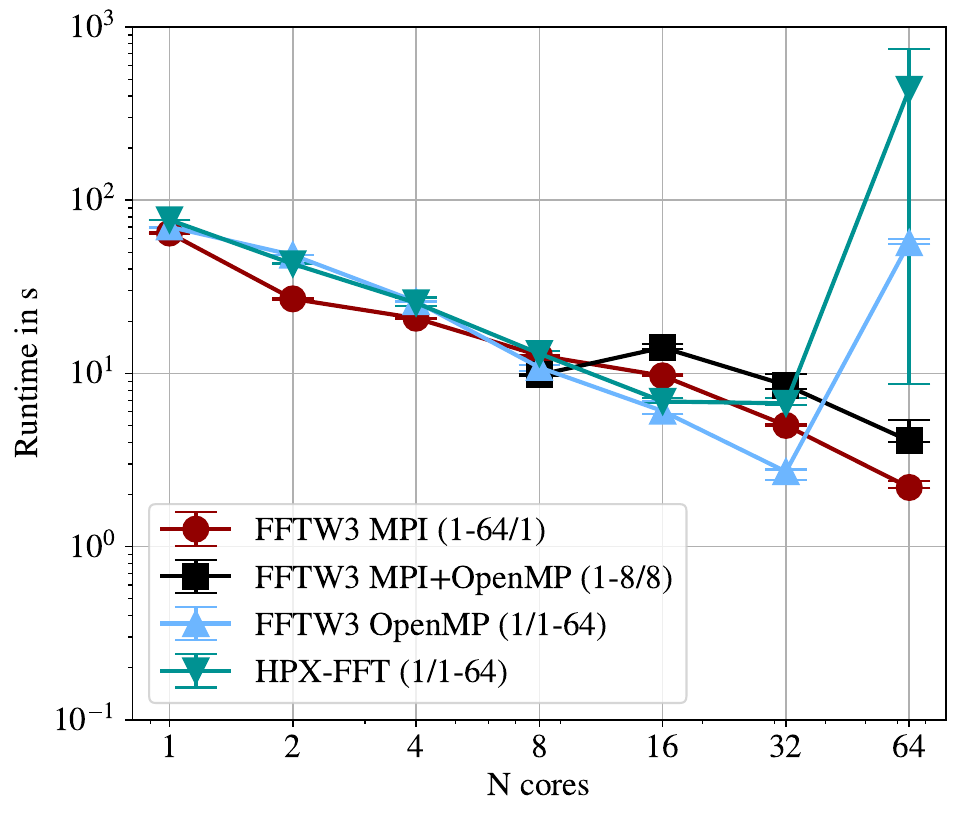}
        \caption{Strong scaling runtimes on up to $64$ RISC-V cores for a $2^{14} \times 2^{14}$ FFT using FFTW with estimated planning. The execution configuration is represented as (\#processes/\#threads).} 
        \label{fig:estimate_riscv}
    \end{minipage}
\end{figure}

The performance of the FFTW backends on the x86-64 machine is visualized in Figure \ref{fig:estimate_x86}. 
For 64 cores, OpenMP performance significantly drops, while MPI scales as expected. 
The HPX-FFT tool shows the best performance and scaling. 
These observations change when we run the same code on the SG2042 RISC-V machine (see Figure \ref{fig:estimate_riscv}). Depending on the backend, we observe an overall performance delta of factor $4.0$ to $8.0$ for parallel FFTW.
Note that FFTW with OpenMP suffers from the same performance drop for 64 cores on RISC-V.

\subsection{Measured planning}

With measured planning, specifically through the creation of a sophisticated parallel plan, the performance of FFTW improves
on the x86-64 system (see Figure \ref{fig:estimate_x86}). 
Compared to estimated planning, HPX-FFT and the MPI backend of FFTW show an average speedup of $1.15$ and $1.27$ respectively.
The OpenMP backend shows an average speedup of $3.12$, including superior scaling for 64 cores. 
Since HPX-FFT and the MPI backend employ similar parallelization strategies, performance gains are limited to improvements in the underlying 1D FFTs. 
In contrast, the OpenMP backend allows FFTW to optimize the entire 2D FFT, resulting in greater overall performance improvements.

On the RISC-V machine, efficient planning utilizing a cycle counter requires the FFTW RISC-V release. 
Although the standard FFTW release can be used on the SG2042 chip, the lack of a cycle counter makes efficient planning impossible.
Alternatively, FFTW can use software timers instead of the cycle counter at the cost of significantly increasing the planning time. 

\begin{figure}
    \centering
    \begin{minipage}[t]{.47\textwidth}
        \centering
        %\raggedright
        \includegraphics[width=\linewidth]{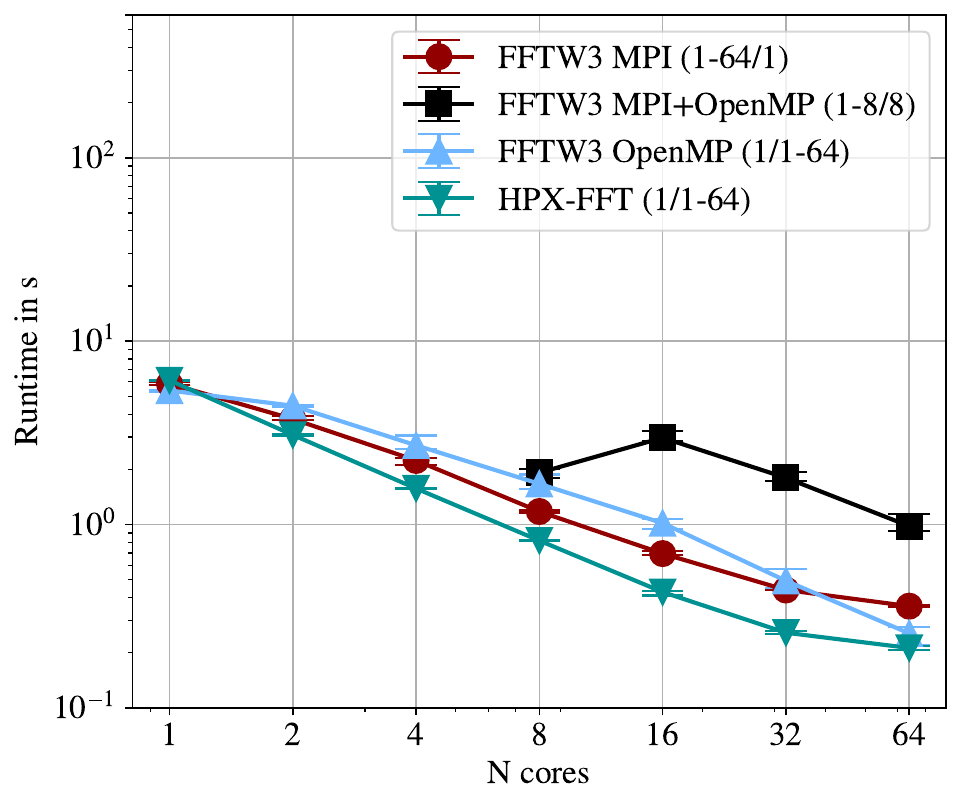}
        \caption{Strong scaling runtimes on up to $64$ x86-64 cores for a $2^{14} \times 2^{14}$ FFT using FFTW with measured planning. The execution configuration is represented as (\#processes/\#threads).}
        \label{fig:measure_x86}
    \end{minipage}\hspace{.05\textwidth}
    \begin{minipage}[t]{.47\textwidth}
        \centering
        %\raggedleft
        \includegraphics[width=\linewidth]{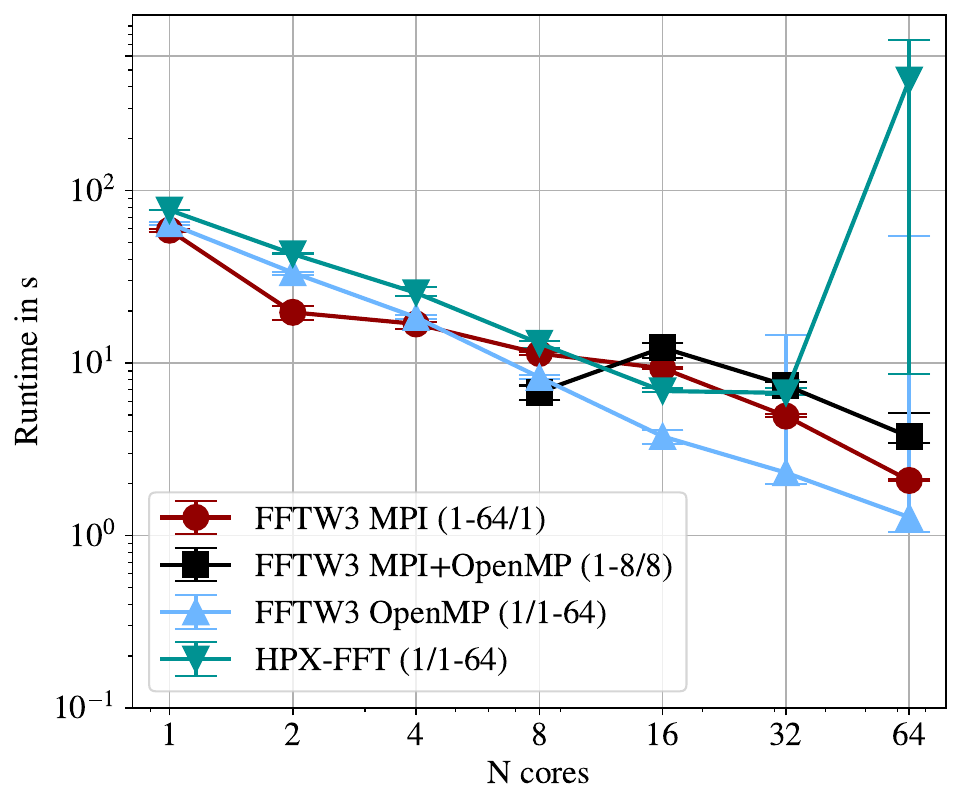}
        \caption{Strong scaling runtimes on up to $64$ RISC-V cores for a $2^{14} \times 2^{14}$ FFT using FFTW with measured planning. The execution configuration is represented as (\#processes/\#threads).} 
        \label{fig:measure_riscv}
    \end{minipage}
\end{figure}

See the performance for FFTW with measured planning on RISC-V in Figure \ref{fig:measure_riscv}. 
Similarly to the x86-64 machine, measured planning resolves the performance drop of the OpenMP backend. 
%Note that in one out of the runs, measured planning resulted in a plan with a runtime similar to estimated planning. 
The MPI backend shows an average speedup of $1.13$, while for OpenMP the speedup is $7.51$ on average.

\section{Conclusion and outlook}\label{sec:conclusion}
In this work, we evaluate the performance of parallel FFT on a recent RISC-V chip using FFTW. 
In a side-by-side comparison of $64$ RISC-V cores against 64 x86-64 cores, we highlight the current performance gap between the two architectures.
We measure a performance delta of around factor eight averaged over all FFT benchmarks conducted in this work. 

Furthermore, we show that optimizing the memory access pattern has significantly less impact on the performance on the SG2042 RISC-V processor. 
Measured planning results in a significant performance boost for the OpenMP backend of FFTW on both chips. 
The MPI backend profits only marginally from planning, but shows consistent scaling. 
The performance advantage of HPX-FFT observed on x86-64 does not translate to RISC-V, as HPX-FFT does not scale beyond 16 cores. Based on these findings, we recommend using FFTW with MPI for parallel FFT on currently available many-core RISC-V systems.

In future work, we plan to consider the power efficiency of the different architectures and compare the performance per watt.
Additionally, we plan to evaluate a broader set of FFT benchmarks, including different problem sizes and configurations with vectorization enabled.
Furthermore, we want to extend the FFT benchmark to ARM-based systems. 
%While RISC-V is making its way into HPC, ARM chips are already part of modern supercomputers, such as the NVIDIA Grace chip or the Fujitsu A64FX chip. 

\section*{Supplementary materials}
The source code, including installation and benchmark scripts, is available at \href{https://doi.org/10.18419/DARUS-5056}{DaRUS}\footnote{\url{https://doi.org/10.18419/DARUS-5056} (accessed on 05/16/2025)}. 
Details on the software and compiler versions used are provided in the \lstinline[language=bash]{README.md} file included with the source code.

\bibliographystyle{splncs04}
\bibliography{bibliography}
\end{document}